\documentclass[a4paper]{article}
\usepackage{epsfig}
\usepackage{graphicx}  
\usepackage{color}     
\usepackage{lscape}
\usepackage{color}
\date{\today}
\begin{document}
\title{\Large \bf Non-zero $\theta_{13}$ from the Triangular Ansatz and Leptogenesis}
{\small
\author{H. B. Benaoum \\
Prince Mohammad Bin Fahd University, Al-Khobar 31952, Saudi Arabia  \\
Email: hbenaoum@pmu.edu.sa}
}
\maketitle
\begin{center}
\small{\bf Abstract}\\[3mm]
\end{center}
Recent experiments indicate a departure from the exact Tri-Bimaximal mixing by measuring definitive non-zero value of $\theta_{13}$. Within the framework of type I seesaw mechanism, we reconstruct the triangular Dirac neutrino mass matrix from the $\mu-\tau$ symmetric mass matrix. The deviation from $\mu-\tau$ symmetry is then parametrized by adding dimensionless parameters $y_i$ in the triangular mass matrix. In this parametrization of the neutrino mass matrix, the non-zero value $\theta_{13}$ is controlled by $\Delta y = y_4 - y_6$. 
We also calculate the resulting leptogenesis and show that the triangular texture can generate the observed baryon asymmetry in the universe via leptogenesis scenario.
\begin{minipage}[h]{14.0cm}
\end{minipage}
\vskip 0.3cm \hrule \vskip 0.5cm

\section{Introduction }
The discovery of the neutrino masses and the lepton mixing have played a pivotal role in probing physics beyond the Standard Model (SM). Recent experiments MINOS ~\cite{minos1,minos2}, T2K ~\cite{t2k}, Double CHOOZ ~\cite{chooz}, Daya Bay ~\cite{daya} and RENO ~\cite{reno} have reported definitive non-zero $\theta_{13}$ result. The latest global analysis of neutrino oscillation data yields the following best-fit values with $1 \sigma, 2 \sigma$ and $3 \sigma$ errors for the oscillation parameters ~\cite{valle}.   \\ 
\begin{table}[!h]  
\begin{center}
\begin{tabular}{|c|c|c|c|} 
\hline
\hline  
Parameter & best fit $\pm 1 \sigma$ & $2 \sigma$ & $3 \sigma$ \\ 
\hline
$\Delta m_{21}^2 [ 10^{-5} eV^2 ]$ & $7.62 \pm 0.19$ & $7.27 - 8.01$ & $7.12 - 8.20$ \\ 
\hline
$\Delta m_{31}^2 [ 10^{-3} eV^2 ]$ & $2.53^{+0.08}_{-0.10}$ & $2.34 - 2.69$ & $2.26 - 2.77$ \\       
                                   & $-(2.40^{+0.10}_{-0.07})$ & $-(2.25 - 2.59)$ & $-(2.15 - 2.68)$ \\ 
\hline
$\sin^2 \theta_{12}$               & $0.320^{+0.015}_{-0.017}$ & $0.29 - 0.35$ & $0.27 - 0.37$ \\ 
\hline
$\sin^2 \theta_{23}$               & $0.49^{+0.08}_{-0.05}$    & $0.41 - 0.62$ &   $0.39 - 0.64$    \\
                                   & $0.53^{+0.05}_{-0.07}$    & $0.42 - 0.62$ &  $0.39 - 0.64$      \\
\hline 
$\sin^2 \theta_{13}$               & $0.026^{+0.003}_{-0.004}$ & $0.019 - 0.033$ & $0.015 - 0.036$ \\              
                                   & $0.027^{+0.003}_{-0.004}$  & $0.020 - 0.034$  & $0.016 - 0.037$ \\
\hline
$\delta$                           & $(0.83^{+0.54}_{-0.64}) \pi$ &    $0 - 2 \pi$  &   $0 - 2 \pi$             \\  
                                   &  $0.07 \pi$                  &  $0 - 2 \pi$    &    $0 - 2 \pi$           \\         
\hline 
\end{tabular}
\caption{Global oscillation analysis with best fit for $\Delta m_{21}^2,\Delta m_{31}^2, \sin^2 \theta_{12}, \sin^2 \theta_{23}, \sin^2 \theta_{13}$ and $\delta$ the upper and/or lower corresponds to normal and/or inverted neutrino mass hierarchy. }
\label{table1}  
\end{center}
\end{table}  
~\\
Seesaw mechanism ~\cite{see1}-\cite{see4} gives a natural explanation of the smallness of the masses for light neutrinos and mixing by connecting the tiny neutrino masses to a very heavy right-handed neutrinos masses. \\
According to the seesaw mechanism, lepton number is broken at high energies due to right-handed neutrino Majorana masses, resulting in small left-handed neutrino Majorana masses suppressed by the heavy mass scale. The seesaw mechanism also provides an attractive mechanism for generating the baryon asymmetry of the universe via leptogenesis ~\cite{lepto1}-\cite{lepto8}. \\
 
The seesaw Lagrangian can be written as : 
\begin{eqnarray}
{\cal L} & = &  \bar{l}_R M_l l_L + \bar{N}_R M_D \nu_L + \frac{1}{2} \bar{N}_R M_R N^c_R + h.c.
\end{eqnarray}
where $\nu_L = ( \nu_e, \nu_{\mu}, \nu_{\tau} )^T, l_L = ( e, \mu, \tau )^T$ and $N_R = (N_1, N_2, N_3)$ denotes 
the left-handed (light) neutrinos, the left-handed charged leptons and the right-handed (heavy) Majorana neutrinos, 
respectively. We assume that the mass matrices of both the heavy $M_R = diag ( M_1, M_2, M_3 )$ and the charged lepton 
$M_l$ are diagonal and real matrices. \\
After integrating out the heavy right-handed neutrinos, the symmetric Majorana mass matrix for light neutrinos is : 
\begin{eqnarray}
M_{\nu} & = & - M_D M_R^{-1} M^T_D   .
\end{eqnarray}

The mixing matrix $V = U P_{Maj}$ that diagonalizes $M_{\nu}$ is : 
\begin{eqnarray}
M_{\nu} & = & V M_{\nu}^{diag} V^{T} 
\end{eqnarray}
where $M_{\nu}^{diag} = diag ( m_1, m_2, m_3 )$ and 
\begin{eqnarray} 
U  & = &  {  \left( 
\begin{array}{ccc}
c_{12} c_{13} & s_{12} c_{13} & s_{13} e^{- i \delta} \\
- s_{12} c_{23} - c_{12} s_{23} s_{13} e^{i \delta} & c_{12} c_{23} - s_{12} s_{23} s_{13} e^{i \delta} & s_{23} c_{13} \\
s_{12} s_{23} - c_{12} c_{23} s_{13} e^{i \delta} & - c_{12} s_{23} - s_{12} c_{23} s_{13} e^{i \delta} & c_{23} c_{13} \\  
\end{array} 
\right) } 
\end{eqnarray}
where $c_{ij} = \cos \theta_{ij}, s_{ij} = \sin \theta_{ij}$, $\delta$ is the CP-violating phase 
and $P_{Maj}$ is diagonal matrix consisting of non-trivial Majorana phases. \\

Now any nonsingular matrix can be decomposed into the product of a nonsingular, upper or lower  triangular matrix $T_{L,U}$ and a unitary matrix $W$. In particular, the Dirac matrix $M_D$ can be written as : 
\begin{eqnarray}
M_D & = & W ~T_{L,U}   .
\end{eqnarray} 
It has been noticed that the relevant quantities for leptogenesis are the CP-violating phases in 
$M_D^{\dagger} M_D = T_{L,U}^{\dagger} T_{L,U}$ which means that the unitary matrix $W$ cancels out in the unflavored 
leptogenesis but not in the seesaw formula. Our ansatz consists in taking the unitary matrix $W$ to be identity implying that the Dirac matrix is either lower or upper triangular matrix $M_D = T_{L,U}$. \\ 
Triangular textures have been considered as the simplest form to study since all the unphysical features can be eliminated from the start ~\cite{benaoum1} and \cite{benaoum2}. Leptogenesis with triangular ansatz has been studied by \cite{branco1,falcone1}. \\
 
The paper is organized as follows. In the next section, we consider a lower or upper triangular matrix as Dirac mass matrix. We show how to reconstruct the triangular Dirac matrix from the neutrino symmetric mass matrix for the Tri-Bimaximal (TBM) mixing. In section 3, we discuss the breaking of the triangular mass by adding dimensionless parameters. We next analyze the mass texture by fitting it to the observed neutrino oscillation data. We show that the triangular mass matrix can naturally accommodate the observed neutrino oscillation parameters. In section 4, we calculate the relevant quantities for leptogenesis and estimate the baryon asymmetry that is consistent with the observed value.

\section{Triangular Dirac Matrix for Tri-Bimaximal Mixing }
In the following, we reconstruct the Dirac matrix $M^{0}_D = T^{0}_{L,U}$ analytically from the neutrino symmetric mass matrix elements $m_{ij}$ ( i.e. light neutrinos masses, heavy Majorana masses and the Majorana phases ) for the Tri-Bimaximal mixing pattern $\theta_{23} = \frac{\pi}{4}, \theta_{12} = \sin^{-1} \left(1/\sqrt{3} \right)$ and 
$\theta_{13} = 0$ ~\cite{harrison}. 
The general symmetric matrix that is diagonalized by the Tri-Bimaximal mixing matrix $V^{0}$ can be written as : 
\begin{eqnarray}
M^{0}_{\nu} & = & - T_{L,U}^{0} M_R^{-1} T_{L,U}{^{0}}^T ~=~V^{0} M_{\nu}^{0~diag} V{^{0}}^T \nonumber\\
& = & \left( \begin{array}{ccc} 
a' & b' & b' \\
b' & c' & a' + b' - c' \\
b' & a' + b' - c' & c' \end{array} \right) 
\end{eqnarray}
with the eigenvalues : 
\begin{eqnarray}
m^0_1 & = & a' - b' \nonumber \\
m^0_2 & = & a' + 2 b'  \nonumber \\
m^0_3 & = & 2 c' - a' - b'   .
\end{eqnarray}

By solving the above equation, we get the following Dirac matrix for upper or lower triangular matrix : 
\begin{eqnarray}
T^0_L & = & - i ~
\left( \begin{array}{ccc} 
\sqrt{a'} & 0 & 0 \\
\frac{b'}{\sqrt{a'}} & \sqrt{\frac{a' c' - b'^2}{a'}} & 0 \\
\frac{b'}{\sqrt{a'}} & \frac{a'^2 + a' b' - b'^2 - a' c'}{\sqrt{a' ( a' c' - b'^2)}} & 
\sqrt{\frac{d'}{a' c' - b'^2}} 
\end{array} \right) ~\sqrt{M_R}
~~~~~~\mbox{or} ~~ \nonumber \\
T^0_U & = & i~ 
\left( \begin{array}{ccc} 
\sqrt{\frac{(a' - b')(a' + 2 b')}{a' + b'}} & - b' \sqrt{\frac{2 c' - a' -b'}{c' (a' + b')}} & \frac{b'}{\sqrt{c'}}  \\
0 & - \sqrt{\frac{(a' + b') (2 c' - a' -b')}{c'}}& \frac{a' + b' - c'}{\sqrt{c'}}  \\
0 & 0 & \sqrt{c'}  
\end{array} \right) ~\sqrt{M_R}
\end{eqnarray}
where $d' = det( M_{\nu} )$ is the determinant of the neutrino mass matrix. \\

\section{Deviations from Tri-Bimaximal}
Now we consider deviation from TBM by breaking the triangular neutrino matrix with dimensionless parameters $y_i$ as :
\begin{eqnarray}
T_L & = & - i ~
\left( \begin{array}{ccc} 
\sqrt{a'}  ~y_1  & 0 & 0 \\
\frac{b'}{\sqrt{a'}} ~y_5 & \sqrt{\frac{a' c' - b'^2}{a'}} ~y_4 & 0 \\
\frac{b'}{\sqrt{a'}} ~y_2 & \frac{a'^2 + a' b' - b'^2 - a' c'}{\sqrt{a' ( a' c' - b'^2)}} ~y_3 & 
\sqrt{\frac{d'}{a' c' - b'^2}} ~y_6
\end{array} \right) ~\sqrt{M_R}
~~~~~~\mbox{or} ~~ \nonumber \\
T_U & = & i~ 
\left( \begin{array}{ccc} 
\sqrt{\frac{(a' - b')(a' + 2 b')}{a' + b'}} ~y_1  & - b \sqrt{\frac{2 c' - a' -b'}{c' (a' + b')}} ~y_2 & \frac{b'}{\sqrt{c'}} ~y_3 \\
0 & - \sqrt{\frac{(a' + b') (2 c' - a' -b')}{c'}} ~y_4 & \frac{a' + b' - c'}{\sqrt{c'}} ~y_5 \\
0 & 0 & \sqrt{c'} ~y_6 
\end{array} \right) ~\sqrt{M_R}
\end{eqnarray}
with $a' = a~e^{i \varphi_a}, b' = b~e^{i \varphi_b}$ and $c' = c~e^{i \varphi_c}$. \\

It is interesting to notice that by taking $y_3 = y_2 = y_6$ and $y_5 = y_4$ for the lower triangular and $y_3 = y_2 = y_1$ and $y_5 = y_4$ for upper triangular, the neutrino matrix $M_{\nu}$ will be : 
\begin{eqnarray}
M_{\nu} & = & 
\left( \begin{array}{ccc} 
a' ~y_1 ^2 & b' ~y_1 y_4 & b' ~y_1 y_6 \\
b' ~y_1 y_4 & c' ~y_4^2 & ( a' + b' - c' ) ~y_4 y_6 \\
b' y_1 y_6  & (a' + b' - c') ~y_4 y_6 & c' ~y_6^2 \end{array} \right)  .
\end{eqnarray}

For $y_4 = y_6 = 1$, $M_{\nu}$ has $\mu - \tau$ symmetry leading to $\theta_{13} = 0$ and $\theta_{23} = \frac{\pi}{4}$. 
Moreover, in the limit $y_1,y_4,y_6 \rightarrow 1$, the above matrix gives the TBM angles and mass eigenvalues. \\ 
In particular for $y_1 = y_2 = y_3 = y_4 = y_5 = 1$, the lower or upper triangular mass matrix gives the neutrino matrix $M_{\nu}$ :
\begin{eqnarray}
M_{\nu} & = & 
\left( \begin{array}{ccc} 
a'  & b'  & b' ~y_6 \\
b'  & c'  & ( a' + b' - c' )  ~y_6 \\
b' ~y_6  & (a' + b' - c') ~y_6 & c' ~y_6^2 \end{array} \right)  
\end{eqnarray}
which is the strong scaling Ansatz studied by \cite{rodejohann}. \\

To see how neutrino mass matrix given by $T_{L,U}$ can lead to deviations of neutrino angles from their TBM values, we consider the hermitian matrix 
$H_{\nu} = M_{\nu} M_{\nu}^{\dagger}$, 
\begin{eqnarray}
H_{\nu} & = & \left( \begin{array}{ccc}
y_1^2 A & y_1 y_4 B & y_1 y_6 C   \\
y_1 y_4 B^{\ast} & y_4^2 D & y_4 y_6 E   \\
y_1 y_6 C^{\ast} & y_4 y_6 E^{\ast} & y_6^2 F \end{array} \right)
\end{eqnarray}
where 
\begin{eqnarray}
A & = & y_1^2 a^2 + ( y_4^2 + y_6^2) b^2  \nonumber \\
B & = & y_1^2 a' b'^{\ast} + y_4^2 b' c'^{\ast} + y_6^2 b' ( a'^{\ast} + b'^{\ast} - c'^{\ast} ) \nonumber \\
C & = & y_1^2 a' b'^{\ast} + y_4^2 b' ( a'^{\ast} + b'^{\ast} - c'^{\ast} ) + y_6^2 b' c'^{\ast} \nonumber \\
D & = & y_1^2 b^2 + y_4^2 c^2 + y_6^2 \left|a' + b' -c' \right|^2 \\
E & = & y_1^2 b^2 +  y_4^2 c' ( a'^{\ast} + b'^{\ast} - c'^{\ast} ) +  y_6^2  ( a' + b' - c' ) c'^{\ast} \nonumber \\
F & = & y_1^2 b^2 +  y_4^2 \left|a' + b' -c' \right|^2 + y_6^2 c^2   .
\end{eqnarray}

As a consequence, the hermitian matrix $H_{\nu}$ is diagonalized by the unitary matrix $V$, 
\begin{eqnarray}
H_{\nu} & = & V M_{\nu}^{diag} M_{\nu}^{diag~ \dagger} V^{\dagger}
\end{eqnarray}
where the deviation of the $\theta_{ij}$ from their TBM values is parametrized by three quantities $\epsilon_{ij}~~(i,j = 1,2,3)$ :

\begin{eqnarray}
\theta_{23} = \pm \frac{\pi}{4} + \epsilon_{23}~~~~~~~~~~~~~~\theta_{13} = \epsilon_{13}~~~~~~~~~~~
\theta_{12} = \sin^{-1} \left( \frac{1}{\sqrt{3}} \right) + \epsilon_{12} ~~~~.
\end{eqnarray}

The angle $\theta_{23}$ of the atmospheric mixing is expressed in terms of the parameters $y_i$ as : 
\begin{eqnarray}
\tan \theta_{23} & = & \frac{y_4 Im (B)}{y_6 Im ( C)} \nonumber \\
& = & \frac{y_4}{y_6}~\frac{a ( y_1^2 - y_6^2 ) \sin \varphi_{ab}  - c (y_4^2 - y_6^2) \sin \varphi_{cb}}
{a ( y_1^2 - y_4^2 ) \sin \varphi_{ab}  + c (y_4^2 - y_6^2) \sin \varphi_{cb} }    
\end{eqnarray}
where $\varphi_{ab} = \varphi_a - \varphi_b$ and $\varphi_{cb} = \varphi_c - \varphi_b$ . \\
 
The Dirac CP phase $\delta$ is given by :
\begin{eqnarray}
\tan \delta & = & - \frac{y_4 y_6 ~Im (E)}{c_{23} s_{23} (y_4^2 D - y_6^2 F ) + y_4 y_6 ( c_{23}^2 - s_{23}^2 ) Re (E)} \nonumber \\
& = &  \frac{2 c y_4 y_6 (y_4^2 - y_6^2) \left[ a \sin \varphi_{ab} - b \sin \varphi_{cb}  \right]}{\Delta}
\end{eqnarray}

where 
\begin{eqnarray}
\Delta & =  &
 (y_4^2 - y_6^2) \left[ b^2 y_1^2 + c^2 ( y_4^2+y_6^2) \right] \sin (2 \epsilon_{23} ) \nonumber \\
& + & 2 y_4 y_6 \left[b^2 y_1^2 + 
 c (y_4^2 +y_6^2 ) ( -c + a \sin \varphi_{ab}  + b \cos \varphi_{cb} ) \right] \cos ( 2 \epsilon_{23}) ~~. 
\end{eqnarray}

In this analysis, we consider only the case of $a' \simeq b'$. In the case, we have $Im (C) \simeq - Im (B)$. \\

The atmospheric angle $\theta_{23}$ is : 
\begin{eqnarray}
\tan \theta_{23} & \simeq & -\frac{y_4}{y_6}  
\end{eqnarray}

and the deviation $\epsilon_{23}$ of the atmospheric mixing will be :
\begin{eqnarray}
\tan \epsilon_{23} & \simeq & - \frac{y_4 - y_6}{y_4 + y_6}   ~~.
\end{eqnarray}
We see that the deviation $\epsilon_{23}$ vanishes for $y_4 = y_6$ as it should be. \\

The Dirac CP phase $\delta$ becomes : 
\begin{eqnarray}
\tan \delta & \simeq &  \frac{c (y_4^2 + y_6^2) \sin \varphi_{ca} }
{a y_1^2 + c (y_4^2 + y_6^2) \cos \varphi_{ca} }  ~~.
\end{eqnarray}

A useful measure of CP violation is given by ~\cite{jarlskog}: 
\begin{eqnarray}
J_{CP} & = & \frac{1}{8} \sin 2 \theta_{12} \sin 2 \theta_{23} \sin \theta_{13} \cos \theta_{13} \sin \delta \nonumber \\
& = & \frac{Im ( H_{\nu~12} H_{\nu~23} H_{\nu~31} )}{\Delta m^2_{21} \Delta m^2_{31} \Delta m^2_{32}} ~~.
\end{eqnarray}
Here we have : 
\begin{eqnarray}
Im ( H_{\nu12} H_{\nu23} H_{\nu31} ) ~\simeq~ 
2 a^3 c y_1^2 y_4^2 y_6^2 (y_4^2 - y_6^2) 
( y_1^2 y_4^2 + y_1^2 y_6^2 + 4 y_4^2 y_6^2 ) ( a^2 + c^2 - 2 a c \cos \varphi_{ca} ) ~\sin \varphi_{ca}  
\end{eqnarray}
which shows that for $y_4 = y_6$, the rephasing invariant $J_{CP}$ vanishes. \\

The reactor mixing angle $\theta_{13}$ is : 
\begin{eqnarray}
\tan 2 \theta_{13} & \simeq & \frac{y_4^2-y_6^2}{a y_1 ( y_1^2+y_4^2+y_6^2)} \sqrt{\frac{a^2 y_1^4 + c (y_4^2+y_6^2) 
\left[c ( y_4^2 + y_6^2) + 2 a y_1^2 \cos \varphi_{ca} \right]}{y_4^2+y_6^2}} ~~~.
\end{eqnarray}

The solar mixing angle $\theta_{12}$ is governed by : 
\begin{eqnarray}
\tan 2 \theta_{12} & \simeq & \frac{ 2 y_1 y_6 \sqrt{ X}}{a (y_1^2 + y_4^2 + y_6^2) (4 y_4^2 y_6^2 -y_1^2 y_4^2 - y_1^2 y_6^2)}
\end{eqnarray}
where 
\begin{eqnarray}
X & = & c (y_1^2 - y_6^2 )^2 (y_4^2 + y_6^2) \left[ c (y_4^2 + y_6^2) +  2 a  y_1^2 \cos \varphi_{ca} \right] \nonumber \\
& + & a^2 \left[ 4 y_4^2 (y_4^2 +y_6^2)^2 (2 y_1^2  +y_4^2 + y_6^2 ) + y_1^4 ( 5 y_4^4 + 2 y_4^2 y_6^2 +y_6^4 ) \right]  
\end{eqnarray}
which for $y_1 = y_4 = y_6$, we get $\tan 2 \theta_{12} = 2 \sqrt{2}$ as expected. \\

Finally, the mass squared differences are given by :
\begin{eqnarray}
\Delta m^2_{21} & \simeq & a^2 (y_1^2 + 2 y_6^2) \left[ y_1^2 + 2 y_6^2 + 4 y_6 (y_4 - y_6) \right] + O ((y_4 -y_6)^2 ) \nonumber \\
\Delta m^2_{31} & \simeq & 4 y_6^3 ( c^2 + a^2 - 2 a c \cos \varphi_{ca} ) \left[y_6 + 2 (y_4 -y_6) \right] + O ((y_4 -y_6)^2 ) ~~~.
\end{eqnarray}

In the numerical analysis, we choose the parameters $y_i, a, c$ and $\varphi_{ca}$ as inputs. With the help of the 
$3 \sigma$ allowed experimental neutrino results, we determine the allowed ranges of these inputs. \\
In Fig.1, we have plotted $\epsilon_{23}$ with respect to $y_6$ and $\Delta y = y_4 - y_6$ and obtain restriction on the parameter space of $y_6$ and $\Delta  y$. 
One can see that $y_6$ and $\Delta y$ are constrained in the ranges $0.97 \leq y_6 \leq 1.01$ and $-0.2 \leq \Delta y \leq 0.2$. \\

\begin{figure}[hbtp]
\centering
\epsfxsize=8cm
\centerline{\epsfbox{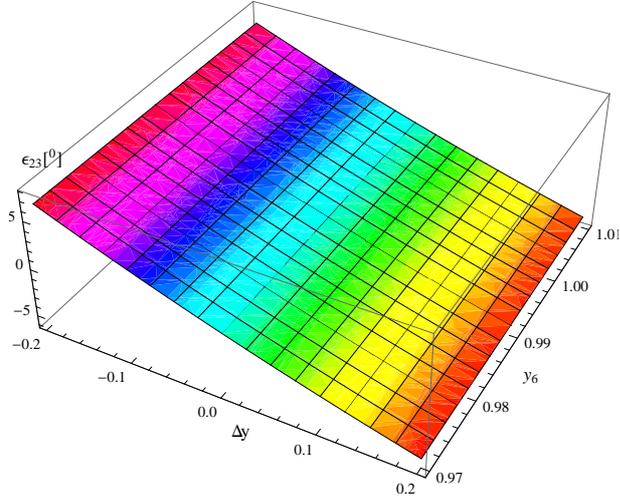}}
\caption{Deviation of the atmospheric angle $\epsilon_{23}$ versus $y_6$ and $\Delta y$.}
\end{figure} 

In the limit $y_1, y_4, y_6 \rightarrow 1$ and $a'\simeq b' \ll c'$, the mass-squared differences $\Delta m^2_{21}$ and 
$\Delta m^2_{31}$ are : 
\begin{eqnarray}
\Delta m^2_{21} & \simeq & 9 a^2 \nonumber \\
\Delta m^2_{31} & \simeq & 4 c^2 .
\end{eqnarray}
Now, using the best fit of the masses-squared differences, we get : 
\begin{eqnarray}
a & \simeq 2.91 \times 10^{-3}~eV \nonumber \\
c & \simeq 2.51 \times 10^{-2}~eV .
\end{eqnarray}

In Fig.2 and Fig.3, we display the exact expressions for the mass-squared differences versus $\Delta y$ and the phase $\varphi_{ca}$ for $y_1 =0.9$ and $y_6 = 0.99$. It can be seen that for these values $\Delta m^2_{21}$ and $\Delta m^2_{31}$ are within $3 \sigma$ ranges. Since $\Delta m^2_{31}$ depends on the $\cos \varphi_{ca}$, Fig. 3 shows two symmetrically regions by varying $\varphi_{ca}$ between $0$ and $2 \pi$. \\

\begin{figure}[hbtp]
\centering
\epsfxsize=8cm
\centerline{\epsfbox{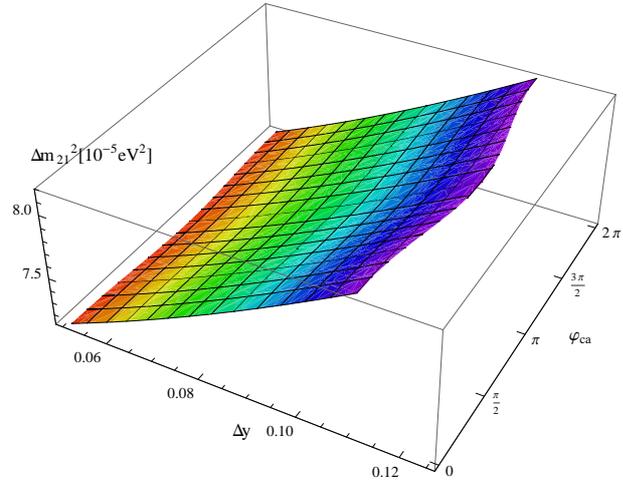}}
\caption{Solar mass squared difference versus $\Delta y$ and $\varphi_{ca}$.}
\end{figure} 

\begin{figure}[hbtp]
\centering
\epsfxsize=8cm
\centerline{\epsfbox{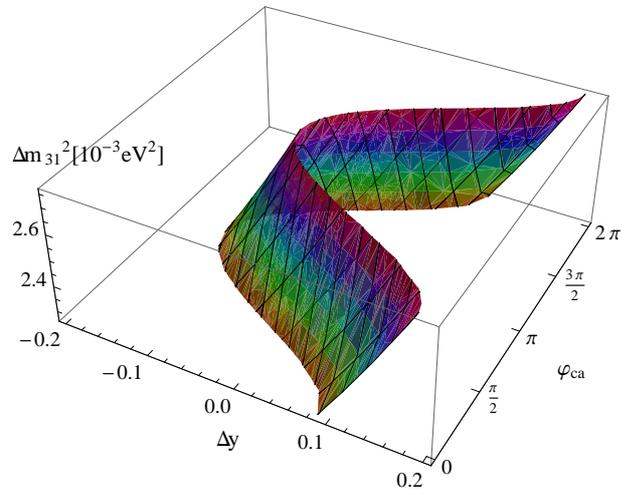}}
  \caption{Atmospheric mass squared difference versus $\Delta y$ and $\varphi_{ca}$.}
\end{figure} 

For $y_1 =0.9$ and $y_6 = 0.99$, we show in Fig.4 and Fig.5 the variation of the reactor $\theta_{13}$ and 
atmospheric $\epsilon_{12}$ angles versus $\Delta y$ and $\varphi_{ca}$. These plots indicate that $\theta_{13}$ and $\epsilon_{12}$ are within $3 \sigma$ ranges. Moreover, since $\theta_{13}$ depends on the $\Delta y$, Fig. 4 shows two symmetrically separated regions within these ranges. \\

\begin{figure}[hbtp]
\centering
\epsfxsize=8cm
\centerline{\epsfbox{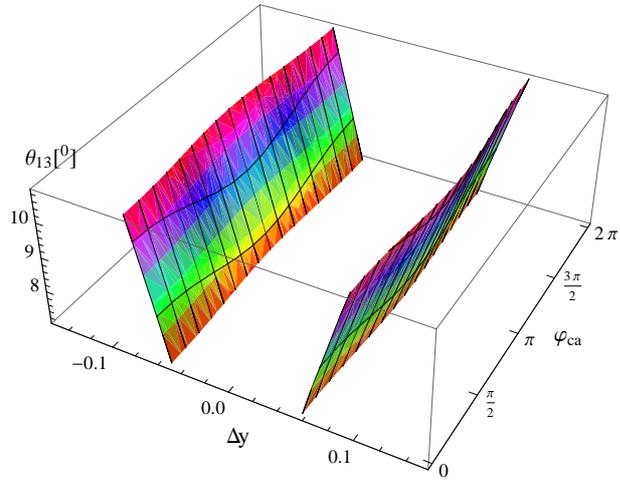}}
\caption{Reactor angle $\theta_{13}$ versus $\Delta y$ and $\varphi_{ca}$.}
\end{figure} 

\begin{figure}[hbtp]
\centering
\epsfxsize=8cm
\centerline{\epsfbox{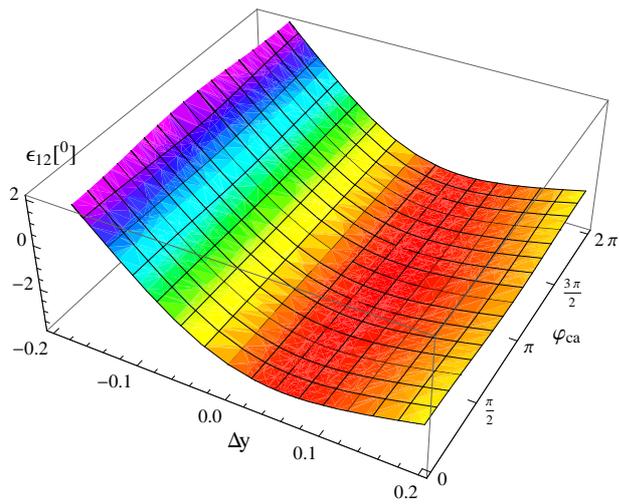}}
  \caption{Deviation of the solar angle $\epsilon_{12}$ versus $\Delta y$ and $\varphi_{ca}$.}
\end{figure} 

\newpage
\section{Leptogenesis}
The CP violating asymmetry is due to the interference of tree level contribution with the one-loop corrections for the decay 
of the ith heavy neutrino ~\cite{lepto1,lepto2} : 
\begin{eqnarray}
\epsilon_i = \frac{\Gamma \left(N_i \rightarrow \phi l^c \right) - \Gamma \left(N_i \rightarrow \phi^{\dagger} l \right)}
{\Gamma \left(N_i \rightarrow \phi l^c \right) + \Gamma \left(N_i \rightarrow \phi^{\dagger} l \right)} ~~.
\end{eqnarray}
The contribution of the heavier neutrinos is washed out and only the asymmetry generated by the lightest neutrino survives :
\begin{eqnarray}
\epsilon_j & = & \frac{3}{16 v^2} \sum_{i \neq j} 
\frac{Im ( M_D^{\dagger} M_D )^2_{ji}}{( M_D^{\dagger} M_D )_{jj}} ~
f \left( \frac{M^2_j}{M^2_i} \right) ~~.
\end{eqnarray}
Here $v \simeq 174~GeV$ is the electroweak symmetry breaking scale and the function $f$ stems from the contributions of both self-energy and vertex diagrams :
\begin{eqnarray}
f \left( x \right) & = & \sqrt{x}~\left( \frac{1}{1 - x} +1 - (1- x) ln ( 1 + 1/x ) \right) ~~.
\end{eqnarray}
For hierarchical heavy Majorana neutrinos i.e., $M_1 \ll M_{2,3}$, one has : 
\begin{eqnarray}
f \left( x \right) & \simeq & - \frac{3}{2 \sqrt{x}} ~~.
\end{eqnarray}
The CP asymmetry $\epsilon_1$ gives rise to a lepton  asymmetry $Y_L$ in the universe : 
\begin{eqnarray}
Y_L & = & \frac{n_L - n_{\overline{L}}}{s} ~=~ \frac{\kappa}{g_{\star}} ~\epsilon_1 
\end{eqnarray} 
where $s$ denotes the entropy density, $g_{\star}\simeq 106.75$ is the effective number of relativistic degrees of freedom contributing to the entropy of the early universe and $\kappa$ is the dilution factor that is obtained by solving the Boltzmann equations. To a good approximation,  $\kappa$ is given by ~\cite{nielsen} : 
\begin{eqnarray}
\kappa \simeq \left\{ \begin{array}{cc} 
\frac{1}{2.0 \sqrt{K^2+9}} & 0 \leq K \leq 10 \\
\frac{0.3}{K (\ln K)^{0.6}} &  10 \leq K \leq 10^6 \\
\end{array} \right. 
\end{eqnarray}
where the parameter $K$, which is defined as the ratio of the thermal average of the $N_1$ decay rate and the Hubble parameter at the temperature $T = M_1$, is : 
\begin{eqnarray}
K & = & \frac{M_P}{1.66 \times 8 \pi v^2 \sqrt{g_{\star}}} \frac{(M_D^{\dagger} M_D )_{11}}{M_1} ~~.
\end{eqnarray}
Here $M_P \simeq 1.22 \times 10^{19}~GeV$ is the Planck mass. \\

The produced lepton asymmetry $Y_L$ is converted into a net baryon asymmetry $Y_B$ through the non-perturbative sphaleron processes. The baryon asymmetry is obtained : 
\begin{eqnarray}
Y_B & = & \frac{\xi}{\xi -1} ~Y_L ~~~~~~~~~~~ \mbox{with}~~~\xi = \frac{8 N_f + 4 N_H}{22 N_f + 13 N_H}
\end{eqnarray}
where in the standard model $N_f = 3$ and $N_H = 1$ are the number of fermion families and complex Higgs doublets respectively. Typically, one gets : 
\begin{eqnarray}
Y_B \simeq - \frac{28}{79} ~Y_L ~~~.
\end{eqnarray}

For the lower triangular Dirac neutrino mass matrix with $a \simeq b \ll c$, the CP asymmetry is : 
\begin{eqnarray}
\epsilon_1 & \simeq & - \frac{3~c M_1 (y_4^2 - y_6^2)^2}{16 \pi v^2 (y_1^2 + y_4^2 + y_6^2)} ~\sin \varphi_{ca} ~~.
\end{eqnarray}

The value of $K$ for the lower triangular texture is expressed as : 
\begin{eqnarray}
K \simeq \frac{a}{10^{-3}} \left( y_1^2 + y_4^2 + y_6^2 \right) 
\end{eqnarray}
which is independent of $M_1$ and is approximately equal to $11.3$ . \\

The baryon asymmetry generated by leptogenesis is : 
\begin{eqnarray}
\eta_B \simeq \frac{3.9 \times 10^{-6} ~( y_4^2 - y_6^2 )^2 ~M_1}
{(y_1^2 + y_4^2 +y_6^2 )^2 \left( 1.3 + \ln (y_1^2 + y_4^2 +y_6^2 ) \right)^{0.6} } ~\sin \varphi_{ca} ~~.
\end{eqnarray}
We see that for an exact $\mu-\tau$ symmetry, the baryon asymmetry $\eta_B$ vanishes \cite{nasri}-\cite{king}. \\

We use $\eta_B = (6.19 \pm 0.15) \times 10^{-10}$ as the upper and lower bound of the baryon-photon ratio from the WMPA observation ~\cite{komatsu}. Our result is shown in Fig.6 with $M_1 = O ( 10^{13} )~GeV$, where the two horizontal lines represents the WMPA allowed bounds. We can see that the triangular Dirac mass matrix provides a baryon asymmetry consistent with the current observations. \\

\begin{figure}[hbtp]
\centering
\epsfxsize=8cm
\centerline{\epsfbox{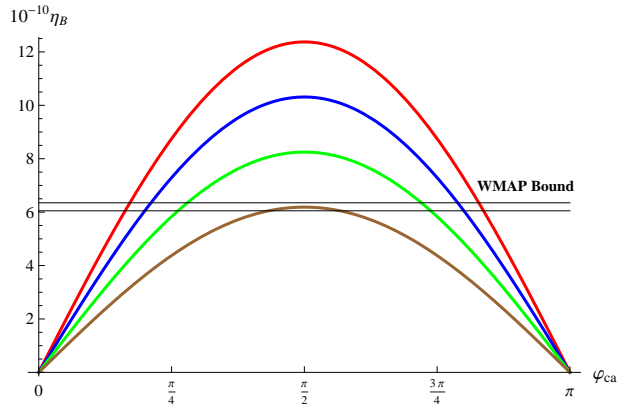}}
\caption{Baryon asymmetry $\eta_B$ versus $\varphi_{ca}$ for $M_1 = \left\{1.5,2,2.5,3 \right\} \times 10^{13}~GeV$ and $\Delta y =0.08$.}
\end{figure} 

\section{Conclusion}
In this paper, we have reconstructed lower and triangular textures of the neutrino Dirac mass matrix that can be obtained via type I seesaw mechanism from the neutrino symmetric mass matrix for the TBM mixing pattern. \\
The recent experiments have reported definitive non-zero value for the reactor mixing angle. Therefore, we have considered deviation of the triangular ansatz from TBM by adding dimensionless parameters. We have studied the phenomenological implications of these triangular textures and have obtained interesting results for the mixing angles and mass-squared 
differences. \\
Furthermore, we have shown that the breaking of the triangular mass matrix from the exact TBM gives a non-zero baryon asymmetry via leptogenesis consistent with the current observations.

\section*{Acknowledgments}
I would like to thank S. Nasri for reading the manuscript and useful comments.

\end{document}